\documentstyle[12pt,epsfig]{article}

\def\gsim{\lower0.5ex\hbox{$\:\buildrel >\over\sim\:$}}
\def\lsim{\lower0.5ex\hbox{$\:\buildrel <\over\sim\:$}}

\def\missET {{\not\!\! E_T}}

\begin{document}
\baselineskip=18pt

\begin{center}
{\large\bf Collider signals of a composite Higgs in the Standard Model with four generations}
\end{center}

\vspace*{0.5 in}

\renewcommand{\thefootnote}{\fnsymbol{footnote}}
\centerline{Shaouly Bar-Shalom$^{a}$\footnote{Electronic address: shaouly@physics.technion.ac.il},
Gad Eilam$^{a}$\footnote{Electronic address: eilam@physics.technion.ac.il},
Amarjit Soni$^b$\footnote{Electronic address: soni@bnl.gov}}

\vspace{0.1 in}

\centerline{\it $^a$Physics Department, Technion-Institute of Technology, Haifa 32000, Israel}
\centerline{\it $^b$Theory Group, Brookhaven National Laboratory, Upton, NY 11973, USA}

\renewcommand{\thefootnote}{\arabic{footnote}}
\setcounter{footnote}{0}
\vspace*{0.5 in}

\begin{abstract}
Recent fits of electroweak precision data to the Standard Model (SM) with a 4th
sequential family (SM4) point to a possible ``three-prong composite solution":
(1) the Higgs mass is at the TeV-scale, (2) the masses of the 4th family quarks
$t^\prime,b^\prime$ are of ${\cal O}(500)$ GeV and (3)
the mixing angle between the 4th and 3rd generation quarks
is of the order of the Cabibbo angle, $\theta_{34} \sim {\cal O}(0.1)$.
Such a manifestation of the SM4 is of particular interest as it
may suggest that
the Higgs is a composite state,
predominantly of the 4th generation heavy
quarks.
Motivated by the above, we show that the three-prong composite solution to the SM4
can have interesting new implications for Higgs phenomenology.
For example, the Higgs can decay
to a single heavy 4th generation quark
via the 3-body decays (through an off-shell $t^\prime$ or $b^\prime$)
$H \to \bar t^\prime t^{\prime \star} \to \bar t^\prime b W^+$ and
$H \to \bar b^\prime b^{\prime \star} \to \bar b^\prime t W^-$.
These flavor diagonal decays can
be dramatically enhanced at the LHC (by several orders of magnitudes)
due to the large width effects of the resonating heavy Higgs
in the processes
$gg \to H \to \bar t^\prime t^{\prime \star} \to \bar t^\prime b W^+$ and
$gg \to H \to \bar b^\prime b^{\prime \star} \to \bar b^\prime t W^-$,
thus yielding a viable signal above the corresponding continuum
QCD production rates.
In addition, the Higgs can decay to a single $t^\prime$ and $b^\prime$
in the loop-generated flavor changing (FC) channels
$H \to b^\prime \bar b ,t^\prime \bar t $.
These FC decays are essentially ``GIM-free" and can, therefore, have
branching ratios as large as $10^{-4} - 10^{-3}$.
\end{abstract}

\newpage

\section{Introduction}

The SM4 is one of the simplest new physics scenarios, in which
the SM is enlarged by a complete sequential 4th family of chiral matter:
a new $(t^\prime, b^\prime)$ and $(\nu^\prime,\ell^\prime)$ heavy doublets in the quark
and lepton sectors, respectively (for reviews see \cite{sher,hou2009-rev,SM4proc}).
For many years, since the LEPI measurement of the Z-width and up to the early 2000's,
the SM4 was thought to be ruled out by experimental data \cite{PDG2008};
however, this interpretation no longer appears to be correct. Indeed recent fits of the SM4
to precision EW observables \cite{kribs,chan,vysotsky} and to flavor data
\cite{bobro,hou2006,gadinew,soniCP}
have clarified that the SM4 is not in disagreement with the present
data. According to these recent studies, the masses
of the 4th generation quark doublet may lie in the range
$300-700$ GeV and should be slightly spilt, i.e., $m_{t^\prime} - m_{b^\prime} \sim 45 -75$ GeV
(consistent also with the current CDF limits on the 4th generation quark masses:
$m_{t^\prime} \gsim 310$ GeV \cite{newlimittp} and  $m_{b^\prime} \gsim 340$ GeV \cite{newlimitbp}).
In turn, in order to be consistent with electroweak (EW) precision data, this
favors a heavier Higgs, thereby, completely removing the tension
between the LEPII bound $m_H \gsim 115$ GeV and
the central (best fitted) Higgs mass value $m_H \sim 85$ GeV obtained
from a fit of the SM to the data.
This effect is mainly due to an interesting interplay between the contributions
of the 4th family fermions to the oblique parameters S and T, as was noted in
\cite{kribs,vysotsky}.

Indeed, an extremely interesting outcome of these new studies
is that the best fitted Higgs mass in the SM4 can be driven up to its unitarity bound,
$m_H \lsim 1$ TeV, if the mixing angle between the 4th and 3rd generation quarks,
$\theta_{34}$, is of the size of the Cabibbo angle (i.e., $\theta_{34} \sim {\cal O}(0.1)$)
and if $m_{t^\prime},~m_{b^\prime} \sim {\cal O}(500)$ GeV \cite{chan}.
The combination of a heavy ${\cal O}(500)$ GeV 4th generation doublet with a TeV-scale
Higgs might be a hint that the Higgs is not a fundamental particle but is, rather,
a composite state, primarily, of the heavy 4th generation quarks
\cite{sher,bardeen,soni-composite,recentcomposite,recentcomposite2}.
In particular, within the SM4, such a heavy Higgs and heavy 4th generation doublet can drive
the Landau pole down to the TeV-scale, in which case, the simplest compositeness condition
(used in top-condensate type models) which relates the masses of the heavy fermions
to that of the Higgs, suggests that
$m_H \sim \sqrt{2} m_{Q}$, where $Q$ is the heavy fermion that forms the composite state
\cite{sher,soni-composite}. Another interesting possibility which was recently
raised in \cite{recentcomposite}
is that if the masses of $t^\prime,b^\prime$ are sufficiently heavy, then
the Higgs quartic and 4th generation Yukawa couplings can have a fixed point at
a scale of several TeV (rather than a Landau pole),
around which the 4th generation heavy quarks can form the composite state.

These recent theoretical developments and experimental indications in favor of
the SM4 with
a heavy Higgs and a heavy 4th generation doublet,
as well as the possibility of an alternative non-SUSY
solution to the hierarchy
problem (see \cite{recentcomposite}), lead us to study what
we henceforward name the ``three-prong composite solution" of the SM4:
\begin{eqnarray}
&m_H \lsim 1 ~{\rm TeV}& \nonumber \\
\nearrow &~~~& \searrow \nonumber \\
\theta_{34} \sim {\cal O}(0.1) &\longleftarrow&
m_{t^\prime,b^\prime} \sim {\cal O}(500)~ {\rm GeV} \nonumber
\end{eqnarray}
which addresses the composite Higgs scenario.

We will show that this three-prong composite solution of the SM4
can have interesting new implications on Higgs phenomenology at the LHC.
In particular, we will focus on the case of
$600 ~ {\rm GeV} \lsim m_H \lsim 1000~ {\rm GeV}$
and  $400 ~ {\rm GeV} \lsim m_{t^\prime},m_{b^\prime} \lsim 600~ {\rm GeV}$
so that $ m_{t^\prime},~m_{b^\prime} < m_H < 2 m_{t^\prime},~ 2 m_{b^\prime}$.\footnote{
Our study complements the case $m_H \lsim 500$ GeV with $m_{t^\prime},~m_{b^\prime} \lsim 400$ GeV
which was discussed in Ref.\cite{kribs}.}
This range of Higgs and 4th generation quark masses
opens up the interesting possibilities of a heavy (composite) Higgs decaying to a single 4th generation
heavy quark
via an off shell $t^\prime$ or $b^\prime$ in the flavor diagonal channels:
\begin{eqnarray}
H \to \bar t^\prime t^{\prime \star} \to \bar t^\prime b W^+  + h.c. ~~ \& ~~
H \to \bar b^\prime b^{\prime \star} \to \bar b^\prime t W^-  +h.c. ~.
\label{3bdecay}
\end{eqnarray}

\noindent or via the 1-loop FC channels:

\begin{eqnarray}
H \to t^\prime \bar t + h.c. ~~ \& ~~ H \to b^\prime \bar b + h.c. \label{decays} ~,
\label{FC}
\end{eqnarray}

As we will show below, the rates for the 3-body decays in (\ref{3bdecay}), which,
within the three-prong composite solution to the SM4, are the only probe of the Higgs Yukawa couplings
to the 4th family quarks,
can be dramatically enhanced due to the large width effects in the heavy Higgs resonance
production processes
\begin{eqnarray}
gg \to H \to \bar t^\prime t^{\prime \star} \to \bar t^\prime b W^+ + h.c. ~~ \& ~~
gg \to H \to \bar b^\prime b^{\prime \star} \to \bar b^\prime t W^- +h.c. ~,
\label{resonance}
\end{eqnarray}
\noindent and can reach a detectable level above the QCD production rate.

In addition, the FC decays in (\ref{FC}) can have branching ratios (BR) as large as
$10^{-4} - 10^{-3}$, owing to the rather large $\theta_{34} \sim {\cal O}(0.1)$
and to the large $t^\prime$ and $b^\prime$ masses which render these
decays essentially GIM-free.

Throughout our analysis to follow we set the masses of the 4th generation leptons
(which enter in the calculation of the total Higgs width) to
$m_{\ell^\prime}=250$ GeV and $m_{\nu^\prime}=200$ GeV. We furthermore assume that
$V_{tb}=V_{t^\prime b^\prime}=1$ and that
the mixing of the 4th generation quarks with the first two generations ones is
much smaller than $\theta_{34}$, in particular, that
$\theta_{34} \equiv V_{t b^\prime} = V_{t^\prime b} >>
V_{t^\prime d},~ V_{t^\prime s},~V_{u b^\prime},~ V_{c b^\prime}$, consistent with current
experimental constraints \cite{kribs,bobro,gadinew}.

\section{Heavy Higgs production at the LHC}

\begin{figure}[htb]
\begin{center}
\epsfig{file=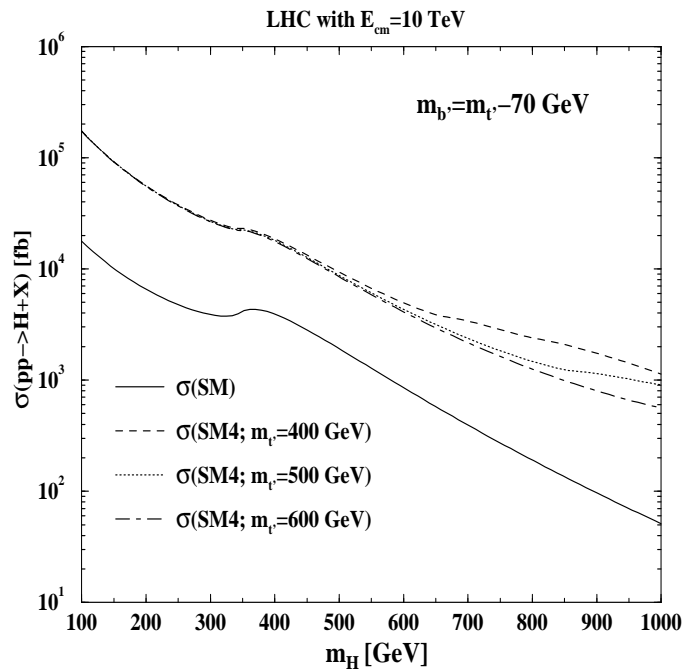,height=9cm,width=9cm,angle=0}
\caption{\emph{$\sigma(pp \to H+X)$ from the $gg \to H$ hard process at the LHC
with a c.m. energy of 10 TeV, in the SM
(solid curve) and in the SM4 for several values of $(m_{t^\prime},m_{b^\prime}$),
as a function of the Higgs mass. We take $K=1.5$ for the K-factor to account
for the QCD corrections to the gluon-gluon fusion cross-section \cite{kfactor}.}}
\label{fig1}
\end{center}
\end{figure}

The Higgs production rate at the LHC is known to increase by almost an order of magnitude in
the SM4, depending on the Higgs and 4th family quark masses \cite{kribs,gunion}.
This is manifest in the
gluon-gluon fusion hard process $gg \to H$ and is caused by the enhancement of the ggH 1-loop
coupling due to the extra heavy 4th family quarks that run in the loop.
This enhancement comes in handy in particular for the production of a TeV-scale Higgs, where
the SM rate might be marginal. In Fig.~\ref{fig1} we plot the gluon-gluon fusion cross-section
convoluted with the PDF for $p p \to H+X$ in the SM and in the SM4 for three
values of the $t^\prime$ mass, $m_{t^\prime} = 400,~ 500$ and $600$ GeV.
Here and henceforward we always set $m_{b^\prime} = m_{t^\prime} - 70$ GeV,
which is the value of $m_{b^\prime}$ appropriate
for Higgs masses $m_H \gsim 600$ GeV.\footnote{Note that according to
Eq.~\ref{split} one has
$65 ~ {\rm GeV} \lsim m_{t^\prime} - m_{b^\prime} \lsim 72~{\rm GeV}$
for $500~{\rm GeV} \lsim m_H \lsim 1000~{\rm GeV}$,
which is the range of Higgs masses of interest in this work.}
That is,
in order for the SM4 to be consistent with EW
precision data the mass splitting in the 4th family quarks is
constrained by \cite{kribs,cont2}:
\begin{eqnarray}
m_{t^\prime} - m_{b^\prime} \approx
\left( 1 + \frac{ \ln\frac{m_H}{115~{\rm GeV}} }{5} \right) \times 50~{\rm GeV} ~.
\label{split}
\end{eqnarray}

As can be seen in Fig.~\ref{fig1}, for $m_H \sim 800$ GeV,
the gluon-fusion cross-section increases from ${\cal O}(100)$ fb
in the SM to ${\cal O}(1000)$ fb in the SM4, thus expecting the LHC to produce about
$10^5$ heavy Higgs with a luminosity of ${\cal O}(100)$ inverse fb.

Note that, when $m_H < 2 m_{t^\prime},2 m_{b^\prime}$, the decay pattern of the
SM4 heavy Higgs
is similar to that of the SM Higgs, up to its possible decays to the 4th family leptons.
In particular, the leading decay
channels of the heavy SM4 Higgs remain $H \to ZZ,WW,t \bar t$. Thus, as was recently noted in \cite{kribs},
the expected enhancement in the gluon-fusion production channel implies that the ``golden mode"
$H \to ZZ \to 4 \mu$ holds up as a useful Higgs discovery channel
throughout the Higgs mass range.

\section{Flavor conserving heavy Higgs decays to a single 4th generation quark \label{sec3}}

Within the three-prong composite solution of the SM4, where
$m_{t^\prime} < m_H < 2 m_{t^\prime}$ and $ m_{b^\prime} < m_H < 2 m_{b^\prime}$, the
2-body Higgs decays $H \to t^\prime \bar t^\prime,~b^\prime \bar b^\prime$ are
kinematically forbidden. In this case, the only probe
of the Higgs Yukawa couplings to the 4th family quarks
are the Higgs 3-body decays
in (\ref{3bdecay}) which proceed through an off-shell heavy 4th family quark.

Here we are interested in the case of vanishing mixings
between the 4th generation quarks with the 1st two generation ones, in which
case $BR(t^\prime \to b W),~BR(b^\prime \to t W) \sim {\cal O}(1)$,
so that the flavor diagonal 3-body Higgs decays in (\ref{3bdecay}) will
lead to:
\begin{eqnarray}
BR(H \to t^\prime \bar t^{\prime \star}) &\sim& BR \left( H \to t^\prime \bar b W^- \to
(b W^+)_{t^\prime} \bar b W^- \right)~, \nonumber \\
BR(H \to b^\prime \bar b^{\prime \star}) &\sim& BR \left( H \to b^\prime \bar t W^+ \to (t W^-)_{b^\prime} \bar t W^+ \right) ~.
\label{3bsignature}
\end{eqnarray}
and the charged conjugate channels.

Note in particular the signature of the $b^\prime$ Yukawa coupling in (\ref{3bsignature}):
$t \bar t  W^-  W^+ \to b \bar b W^+ W^+ W^- W^-$,
which, with the appropriate kinematical cuts, is expected to be rather distinct, e.g.,
leading to a signal of same-sign leptons + missing energy accompanied by 2 b-jets and 4 light-jets:
$H \to b^\prime \bar b^{\prime \star} \to \ell^+ \ell^+ b b jjjj + \missET$ (and the charged conjugate
signal).

Naively one may expect these 3-body decays to have very small rates, with BR's
at the level of $10^{-6}-10^{-5}$ when
$m_H < 2 m_{t^\prime}, 2 m_{b^\prime}$. However, as we will show below,
these decays can be dramatically enhanced due to finite width effects of the
heavy Higgs. Indeed, the non-zero width of heavy particles is known to
cause substantial enhancement when these particles appear in the final state
of a decay which occurs just around its kinematical threshold
\cite{widtheffects}. Alternatively, when the decaying heavy particle emerges
as an intermediate state the effects of its width close to threshold
is usually handled with the Breit-Wigner prescription.
In this case, special attention is required for the computation of the overall
resonant production and subsequent decay of the heavy particle.

Here we consider the leading resonance production of the heavy
Higgs via the gluon-fusion process (which is further enhanced in the SM4, see previous section),
followed by its decays $H \to F$ as in (\ref{3bsignature}).
Using the relativistic Breit-Wigner resonance formula, we estimate
the corresponding hard cross-sections by:
\begin{eqnarray}
\hat \sigma_H(\hat s) = \hat \sigma_{gg}(\hat s) \cdot \hat s \cdot \hat{BW}(\hat s) \cdot BR(H \to F)(\hat s)~,
\end{eqnarray}
with
\begin{eqnarray}
\hat \sigma_{gg}(\hat s) = \frac{\pi^2}{8 m_H^3} \Gamma(H\to gg)(\hat s)
\end{eqnarray}
and $\hat{BW}$ is the normalized Breit-Wigner function:
\begin{eqnarray}
\hat{BW}(\hat s) = \frac{m_H \Gamma_H/\pi}{\left[ (\hat s - m_H^2)^2 + m_H^2 \Gamma_H^2\right]}~,
\end{eqnarray}
where, in order to incorporate the appropriate kinematic dependencies away from the resonance,
 we replace $m_H \Gamma_H \to \sqrt{\hat s} \Gamma_H(\hat s)$ and set
 $\Gamma_H(\hat s)= \sqrt{\hat s} \Gamma_H/m_H$ ($\Gamma_H$ being the total Higgs width) \cite{PDG2008}.
We thus get (for the hard process $gg \to H \to F$):
\begin{eqnarray}
\hat \sigma_H(\hat s) = \frac{\pi \left(\hat s/m_H^2 \right) \Gamma(H\to gg)(\hat s)
\Gamma(H\to F)(\hat s)}{8 \left[ (\hat s - m_H^2)^2 + (\hat s \Gamma_H/m_H)^2 \right]} ~,
\label{BWformula}
\end{eqnarray}
which, after being convoluted with the gluons distribution functions (PDF's):
\begin{eqnarray}
\frac{dL}{d\tau}(\hat s = \tau s)  \equiv \int_{\tau}^{1} \frac{dx}{x} g(x,\hat s) g(\frac{\tau}{x},\hat s) ~,
\end{eqnarray}
gives:
\begin{eqnarray}
\sigma_H = \int_{\tau_{min}}^{\tau_{max}} \frac{dL}{d\tau} \hat\sigma_H d\tau ~,
\end{eqnarray}

In the Narrow Width Approximation (NWA), where $\hat BW(\hat s) \approx \delta(\hat s -m_H^2)$, we obtain:
\begin{eqnarray}
\sigma_H (NWA) = \hat \sigma_{gg} \tau_H \frac{dL}{d\tau}  BR(H\to F)~,
\label{NWA}
\end{eqnarray}
where $\tau_H = m_H^2/s$ and all the other terms in (\ref{NWA}) are evaluated at $\hat s =m_H^2$.

Using the Breit-Wigner formula in (\ref{BWformula}), we have estimated the width effects of the heavy Higgs
on the cross-sections for
producing a single 4th family quark through the 3-body Higgs decays:
\begin{eqnarray}
\sigma_H(t^\prime b W) &\equiv& \sigma( pp \to H +X \to \bar t^\prime t^{\prime \star} + h.c. +X \to
\bar t^\prime b W^+ +h.c. + X )
~,\nonumber \\
\sigma_H(b^\prime t W) &\equiv& \sigma( pp \to H +X \to \bar b^\prime b^{\prime \star} +h.c. +X \to
\bar b^\prime t W^- +h.c. + X ) ~,
\label{resCSX}
\end{eqnarray}
by integrating the corresponding hard cross-sections over some finite range around the resonance,
depending on the Higgs mass and width.
\begin{figure}[htb]
\begin{center}
\epsfig{file=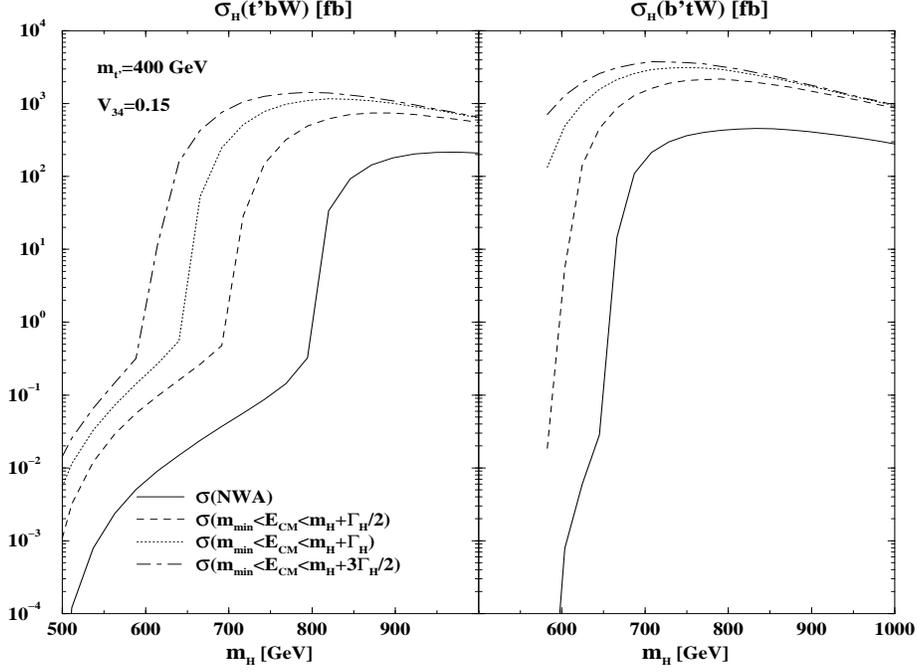,height=12cm,width=9cm,angle=270}
\caption{\emph{$\sigma_H (t^\prime b W)$ and $\sigma_H (b^\prime t W)$ (see Eq.~\ref{resCSX}),
and the corresponding cross-sections in the NWA (see Eq.~\ref{NWA}),
as a function of $m_H$, for
$m_{t^\prime}=400$ GeV ($m_{b^\prime} = m_{t^\prime} - 70$ GeV) and $\theta_{34} =0.15$.
The hard c.m. energy, $\sqrt{\hat s}$, is
integrated from $m_{min}$ to
$m_H + \delta m$, where $\delta m=\Gamma_H/2$ (dashed line), $\delta m=\Gamma_H$ (dotted line)
and $\delta m= 3 \Gamma_H/2$ (dotted-dashed line), and $m_{min}$ is the largest of either the kinematic threshold
or the value of $m_H-\delta m$.}}
\label{fig3}
\end{center}
\end{figure}
\begin{table}[]
\begin{center}
\begin{tabular}{c|c|c|c|c}
~ & \multicolumn{2}{c|}{$pp \to t^\prime b W + h.c. + X$} & \multicolumn{2}{c}{$pp \to b^\prime t W +h.c. + X$} \\
\hline
  Integrated range& $\sigma_H (t^\prime b W)$ & $\sigma_{QCD} (t^\prime b W)$  & $\sigma_H(b^\prime t W)$  & $\sigma_{QCD}(b^\prime t W)$  \\
~ & [fb] & [fb] & [fb] & [fb] \\
\hline
~ & ~ & ~ & ~ & \\
$700 ~{\rm GeV} \leq \sqrt{\hat s} \leq m_H+ \frac{1}{2}\Gamma_H$ &  0.11 & 1.3 & 96 & 634  \\
~ & ~ & ~ & ~ & \\
$700 ~{\rm GeV} \leq \sqrt{\hat s} \leq m_H+ \Gamma_H$ & 58 & 280 & 504 & 1716 \\
~ & ~ & ~ & ~ &  \\
$700 ~{\rm GeV} \leq \sqrt{\hat s} \leq m_H+ \frac{3}{2}\Gamma_H$ & 274 & 708 & 872 & 2358  \\
~ & ~ & ~ & ~ &  \\
$750 ~{\rm GeV} \leq \sqrt{\hat s} \leq 1050 ~{\rm GeV}$ & 14 & 116 & 354 & 1406 \\
~ & ~ & ~ & ~ &  \\
\hline
$\sigma_H(NWA)$ &  \multicolumn{2}{c|}{$4.8 \cdot 10^{-3}$} & \multicolumn{2}{c}{$5.9 \cdot 10^{-3}$} \\
%
\hline
\end{tabular}
\caption{Cross-sections for the flavor-diagonal single $t^\prime$ and $b^\prime$ production in
the Higgs resonance channels:
$\sigma_H (t^\prime b W)$ and
$\sigma_H (b^\prime t W)$ (see Eq.~\ref{resCSX}) and the corresponding
cross-sections from the QCD continuum: $\sigma_{QCD} (t^\prime b W)$ and $\sigma_{QCD} (b^\prime t W)$.
The cross-sections are for the LHC with a c.m. energy of 10 TeV, for
$m_H=800$ GeV (for which we have
$\Gamma_H \sim 300$ GeV), $m_{t^\prime}=500$ GeV,
$m_{b^\prime}=430$ GeV and $\theta_{34}=0.15$, and are
integrated over several mass ranges of the c.m. energy of the hard
processes as indicated.
The corresponding cross-sections $\sigma_H(NWA)$ in the NWA are also given.}
\label{tab1}
\end{center}
\end{table}

In Fig.~\ref{fig3} we give a sample of our results which show the dramatic
enhancement near the threshold for the 3-body Higgs decays
$H \to \bar t^\prime b W^+,\bar b^\prime t W^-$,
due to the large width of the decaying Higgs when it is produced at resonance at the
LHC. Notice that, due to the large Higgs width, the NWA is
not adequate for estimating these cross-sections (as expected). This can be seen
in the figure for $m_H$ masses sufficiently away from the threshold of
the decay to on-shell pair of heavy 4th generation quarks
$H \to t^\prime \bar t^\prime,b^\prime \bar b^\prime$.
For example, for $m_H =1000 ~ {\rm TeV}$ and  $m_{t^\prime}=400$ GeV,
we find $\sigma_H(BW) / \sigma_H(NWA) \sim 3$, where $\sigma_H(BW)$
is the Breit-Wigner estimate of (\ref{BWformula}) for the corresponding
cross-sections.
This difference between the NWA and the BW approaches indicates
the extent to which the NWA is valid in this case.

We see that around the threshold the enhancement
due to the finite width effect of the resonating Higgs
can reach several orders of magnitudes, elevating these cross-sections to
${\cal O}(1000)$ [fb], which should be within the detectable level at the LHC.
This enhancement is
practically independent of the mixing angle $\theta_{34}$, since
as long as the mixings of the 4th generation with the 1st and the 2nd generation quarks
are much smaller than $\theta_{34}$ the decays $t^\prime \to b W$
and $b^\prime \to t W$ occur with a BR of order one.

In Table \ref{tab1} we evaluate these cross-sections for
$m_H=800$ GeV, $m_{t^\prime}=500$ GeV and we list the potential (leading) background
to these single 4th generation quark signals from
the QCD continuum production.\footnote{$\sigma_{QCD} (t^\prime b W)$ and
$\sigma_{QCD}(b^\prime t W)$ were calculated by means of CalcHep \cite{calchep}.}
We see that with a luminosity
of the ${\cal O}(100)$ inverse fb, one expects $10^4 - 10^5$ such single
$t^\prime$ and $b^\prime$ events with a
(naive) signal to background ratio of $S/B \sim 1/10 - 1/3$ and a corresponding
statistical significance of $S/\sqrt{B} \sim 50 - 200$, depending on the range
of integration of the hard cross-sections. Nonetheless,
the detection of these Higgs decay signals at the LHC is rather challenging.
In particular, after the $t^\prime$ and $b^\prime$ decay (promptly) via $t^\prime \to b W,~b^\prime \to t W$
these signals will give two distinct final state topologies:
\begin{eqnarray}
&&pp \to H \to \bar t^\prime t^{\prime \star} \to  (\bar b W^-)_{t^\prime} b W^+  ~ \left[+ ~h.c. +X \right]
~,\nonumber \\
&&pp \to H \to \bar b^\prime b^{\prime \star}  \to
  (\bar t W^+)_{b^\prime} t W^- \to  (\bar b W^+ W^-)_{b^\prime} b W^+ W^- ~\left[+ ~h.c. +X \right] ~,
\label{topology}
\end{eqnarray}
which lead to large multiplicity events with leptons, jets and missing transverse energy
(the $t^\prime \bar t^\prime$ one resembling the case of $t \bar t$ + jets events in the SM).
The reconstruction of such events will, therefore, require a detailed and rather demanding study of both the
signal and background samples.

For example, it is possible to use subsets of the above signatures which can be distinguishable even without a full reconstruction of the event. In fact, such analysis for $b^\prime$ and $t^\prime$-pair production topologies similar to (\ref{topology}) was already performed by CDF in \cite{newlimittp,newlimitbp}, and may be used
as a starting point for the search of our heavy Higgs resonance signals
$H \to \bar t^\prime t^{\prime \star}, ~\bar b^\prime b^{\prime \star}$ at the LHC.
In particular, for the detection of $b^\prime$ via $b^\prime \bar b^\prime \to t \bar t W^+ W^- \to b \bar b W^+W^+W^-W^-$,
\cite{newlimitbp} required the signature $\ell^\pm \ell^\pm b j \missET$:
two same-charge reconstructed leptons, at least two jets one of them b-tagged and missing transverse
energy (with cuts on the transverse energy and rapidity of the jets/leptons, see \cite{newlimitbp}). This
same-sign leptons signature was found to be particularly sensitive to the new physics signal $b^\prime \bar b^\prime \to t \bar t W^+ W^- \to b \bar b W^+W^+W^-W^-$ (the main background coming
from $t \bar t$ and W+jets production) and was used to obtain a bound on the corresponding $b^\prime$ mass $m_{b^\prime} \gsim 340$ GeV \cite{newlimitbp}.
Such an analysis can be similarly applied to our signal (at the LHC)
$\bar b^\prime b^{\prime \star}  \to
  (\bar t W^+)_{b^\prime} t W^- \to  (\bar b W^+ W^-)_{b^\prime} b W^+ W^- $, in particular since it does not rely on the full reconstruction of the  $b^\prime \bar b^\prime$ system.

In the $t^\prime \bar t^\prime \to b \bar b W^+ W^-$ case CDF required a signature of one lepton, four or more jets and missing transverse energy $\ell +4j + \missET$ \cite{newlimittp}, which allowed a useful discrimination of the $t^\prime \bar t^\prime$ signal from the background (here also the SM background mostly consists of $t \bar t$ and W+jets events as well as multi hadronic jet events from QCD).  An analysis in this spirit might also be useful for searching for our $H \to \bar t^\prime t^{\prime \star} \to  (\bar b W^-)_{t^\prime} b W^+$ signal at the LHC.

However, we should stress that the event topologies and the background problems are expected
to be more serious at the LHC and are likely to represent a serious challenge.
Nonetheless, we are cautiously optimistic that it can be handled.

\pagebreak

\section{Flavor changing Higgs decays to a single 4th generation quark}

In the SM the FC decay $H \to t \bar c$ is severely suppressed by the GIM-mechanism,
leading to an un-observably small $BR_{SM}(H \to t \bar c) \sim 10^{-13}$ \cite{HtotcSM}.
On the other hand, in the SM4, we expect this decay mode and also the
FC decays involving the 3rd and 4th generation quarks, $H \to t^\prime \bar t$ and
$H \to b^\prime \bar b$, to be significantly enhanced
due to the extra heavy 4th generation
quarks in the loops which essentially removes the GIM suppression.
This can be seen by estimating the width for the 1-loop FC decay $H \to U \bar u$
in the limit that only the heaviest down-type quark ($D$) runs in the loop and its contribution
is multiplied by the appropriate GIM-suppression factor (see e.g., \cite{behar}):
\begin{eqnarray}
\Gamma(H \to U \bar u) \sim \left(\frac{|V_{U D}^\star V_{u D}|}{16 \pi^2} \right)^2
\left( \frac{g^2}{4 \pi} \right)^3 \left(\frac{m_D}{m_W} \right)^4 m_H ~.
\label{HtcSM}
\end{eqnarray}

For example, we can use (\ref{HtcSM}) to estimate the ratio between $\Gamma_{SM}(H \to t \bar c)$
(where $D=b$)
and
$\Gamma_{SM4}(H \to t^\prime \bar t)$ (where $D=b^\prime$):
\begin{eqnarray}
\frac{\Gamma_{SM4}(H \to t^\prime \bar t)}{\Gamma_{SM}(H \to t \bar c)} \sim
\left(\frac{m_{b^\prime}^2 |V_{t^\prime b^\prime}^\star V_{t b^\prime}|}{m_b^2 |V_{t b}^\star V_{c b}|} \right)^2 ~.
\label{estimate}
\end{eqnarray}

Thus, taking $m_{b^\prime}=500$ GeV, $V_{t^\prime b^\prime}=V_{t b}=1$ and
$V_{t b^\prime} \equiv \theta_{34} \sim 3 V_{cb} \sim 0.15$,
and assuming that the total width of the heavy Higgs in the SM and in the SM4 is the same
(this roughly holds for $ m_H < 2 m_{b^\prime}, 2m_{t^\prime}$), we expect
$BR_{SM4}(H \to t^\prime \bar t) \sim 10^9 \cdot BR_{SM}(H \to t \bar c) \sim 10^{-4}$,
which, as we will show below, is indeed the case. On the other hand, for the case
of $H \to t \bar c$ in the SM4, we expect (using the above values for the CKM elements
involved and for $m_{b^\prime}$)
$BR_{SM4}(H \to t \bar c) \sim V_{c b^\prime}^2 \cdot 10^9 \cdot BR_{SM}(H \to t \bar c)$,
which at best
gives $BR_{SM4}(H \to t \bar c) \sim 10^6 \cdot BR_{SM}(H \to t \bar c) \sim 10^{-7}$
once $V_{c b^\prime}$ is set to its largest allowed value $V_{c b^\prime} \sim 0.03 - 0.04$,
see e.g. \cite{gadinew}.

This expected enhancement in FC transitions involving the 4th family heavy quarks
of the SM4, has ignited a lot of activity in the past two decades.
However, previous studies of FC effects in the SM4
have assumed that $m_{t^\prime},m_{b^\prime}>m_H$ and, therefore, focused on
the FC decays $b^\prime \to bH,bV$
\cite{hou2006,gadinew,GSH,bpdecays} and $t^\prime \to tH,tV$ \cite{hou2006,gadinew},
where $V=g,\gamma,Z$.
The BR's for these $t^\prime$ and $b^\prime$ FC decays were found to be typically within the range
of $10^{-4} - 10^{-2}$. The authors of \cite{hou2006,gadinew} have also
studied  the (SM-like) FC decays $t \to cH,cV$ within the SM4 and found that, in spite of the many orders
of magnitudes enhancement, these FC top decays remain below the LHC sensitivity, i.e.,
having a BR typically smaller than $10^{-7}$.

Here, motivated by the estimate in (\ref{estimate}),
we wish to study the FC heavy Higgs decays in (\ref{FC}):
$H \to t^\prime \bar t$ and $H \to b^\prime \bar b$. Recall that our working assumption
is that $V_{t^\prime b} \sim V_{t b^\prime} >> V_{t^\prime d},
V_{t^\prime s},V_{u b^\prime},V_{c b^\prime}$ and
$m_{b^\prime} = m_{t^\prime} - 70~ {\rm GeV} \gsim 350$ GeV, in which case
the 4th
generation quarks are expected to decay mainly via $t^\prime \to bW$
and $b^\prime \to t W$. Thus, both FC Higgs decays will lead to the final state
$b \bar b W^+ W^-$, but with different kinematics:
\begin{eqnarray}
H \to t^\prime \bar t + h.c. &\to& (b W^+)_{t^\prime} (\bar b  W^-)_t + h.c. ~, \nonumber \\
H \to b^\prime \bar b + h.c. &\to& (t W^-)_{b^\prime} \bar b + h.c. \to (b W^+ W^-)_{b^\prime} \bar b +h.c.
\label{FCkin} ~.
\end{eqnarray}

\begin{figure}[htb]
\begin{center}
\epsfig{file=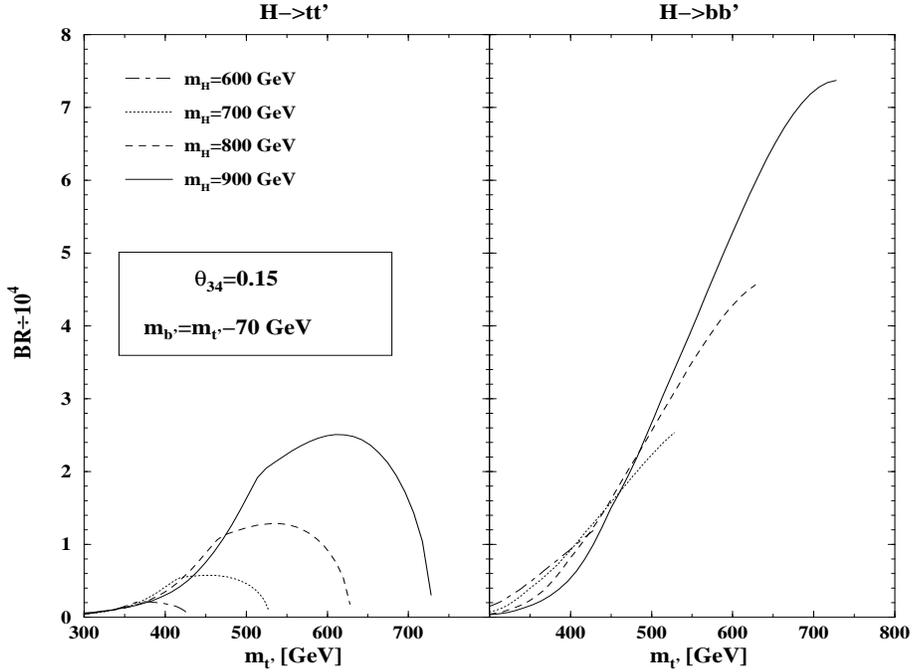,height=12cm,width=9cm,angle=270}
\caption{\emph{$BR(H \to t^\prime \bar t + h.c.)$ (left) and
$BR(H \to b^\prime \bar b + h.c.)$ (right), as a function
of $m_{t^\prime}$ for $\theta_{34} =0.15$,
$m_{b^\prime} = m_{t^\prime} - 70$ GeV and for $m_H = 600,~700,~800$ and $900$ GeV.}}
\label{fig2}
\end{center}
\end{figure}

We base our results below on the explicit analytical expressions that were
given for the decay $H \to b^\prime \bar b$ in \cite{GSH}. In Fig.~\ref{fig2}
we plot the $BR(H \to t^\prime \bar t + h.c.)$ and $BR(H \to b^\prime \bar b + h.c.)$,
as a function of $m_{t^\prime}$, for $\theta_{34}=0.15$ and for Higgs masses
between 600 to 900 GeV. We see, for example, that for $m_{t^\prime} \sim 500$ GeV
these BR's are at the level of ${\rm few}\times 10^{-4}$ if $m_H \sim 800$ GeV
and $\theta_{34} \sim 0.15$, which is the value
of $\theta_{34}$ required in order
for the SM4 to fit EW precision measurements when $m_H \sim 800$ GeV \cite{chan}.
Note that in Fig.~\ref{fig2} we did not consider the finite width effects of the heavy Higgs.
We thus expect the BR's in Fig.~\ref{fig2} to be enhanced
by an additional factor of a few, i.e., reaching $10^{-3}$, when
these FC signals are integrated over some finite range around $m_H$
in the Higgs production process, e.g., integrated
over $m_H - \Gamma_H < \sqrt{\hat s} < m_H + \Gamma_H$ in
$gg \to H \to t^\prime \bar t, b^\prime \bar b$
(see discussion in the previous section).
Given that the LHC with
a luminosity of ${\cal O}(100)$ fb$^{-1}$ will be able to produce about $10^5$
TeV-scale Higgs particles (see Fig.~\ref{fig1}), one naively expects tens
of such events a year, if indeed $\theta_{34} \sim 0.15$.

\pagebreak

\section{Summary and discussion}

We have explored some of the phenomenological implications
for heavy Higgs physics in the SM4 within
what we named the ``three-prong composite solution" to EW precision data,
which accommodates the possibility of a composite Higgs:
(1) $m_H \sim {\cal O}(1)$ TeV, (2) $m_{t^\prime,b^\prime} \sim {\cal O}(500)$ GeV and
(3) $V_{t^\prime b},V_{t b^\prime}$ of the order of the Cabibbo angle, i.e.,
$V_{t^\prime b} \sim V_{t b^\prime} \sim \theta_{34} \sim {\cal O}(0.1)$, and
we have
assumed that $m_H < 2 m_{b^\prime},2 m_{t^\prime}$ in accordance with compositeness.

We focused on
the flavor diagonal  $H \to \bar t^\prime t^{\prime \star} \to \bar t^\prime b W^+$
and $H \to \bar b^\prime b^{\prime \star} \to \bar b^\prime t W^-$ 3-body
decay channels and the flavor changing (one-loop) $H \to t^\prime \bar t, b^\prime \bar b$ decays
of the composite Higgs to a single heavy 4th family quark. We are cautiously optimistic that these signals
can be observed at the LHC, in spite of the
decrease in
the production cross-section for such a heavy Higgs.
In particular, the one-loop FC decay channels are essentially ``GIM-free" and
can, therefore, reach a branching ratio as large as $BR \sim 10^{-4} - 10^{-3}$
if $\theta_{34} \sim {\cal O}(0.1)$,
while the rate of the flavor diagonal channels is dramatically enhanced due
to large width effects of the decaying Higgs - potentially yielding
$10^4 - 10^5$ such events with a luminosity of ${\cal O}(100)$ inverse fb,
which is comparable to the number of events expected
from
the QCD continuum production of $\bar t^\prime t^{\prime \star}$ and
$\bar b^\prime b^{\prime \star}$.

These mechanisms for producing single heavy 4th family fermions
are particularly interesting for the LHC when applied
to the leptonic sector, due to the absence of the
QCD background. As noted in \cite{kribs} (see also \cite{unel}), the Higgs decay
to a pair of on-shell $\nu^\prime$ followed by $\nu^\prime \to \ell W$,
can
yield a rate to a four lepton final state (via
$H \to \nu^\prime \bar\nu^\prime \to 4 \ell + \missET$) which
is comparable to the rate expected from the ``golden mode"
$H \to ZZ \to 4 \ell$, if the mixing between the 4th generation
leptons with the lighter leptons (via
the charged current $U_{i4} \ell^\pm_i \nu^\prime W^\mp$, $i=1,2$ or $3$)
is not exceedingly small.
However, if
$m_H < 2 m_{\nu^\prime},2m_{\ell^\prime}$, then the decays
$H \to \nu^\prime \bar\nu^\prime, \ell^\prime \bar\ell^\prime$ are
kinematically not open and, as in the quark sector, the Higgs Yukawa
couplings to the 4th generation leptons can be probed only through
its decays to a single $\nu^\prime$ and $\ell^\prime$. For instance,
applying our results above to the leptonic channels, e.g., taking
$m_H \sim 800$ GeV and $m_{\nu^\prime} \sim 500$ GeV
we expect the BR to the flavor diagonal 3-body decay
$H \to \nu^\prime \bar\nu^{\prime \star}$, followed by
$\nu^\prime \to \ell W$, to be dramatically enhanced
due to the large width effects around the heavy Higgs resonance in
$gg \to H \to \nu^\prime \ell W$ (independent of the mixing angle
$U_{i4}$, see section \ref{sec3}), thus yielding
about $10^4$ $H \to \nu^\prime \ell W \to (\ell W)_{\nu^\prime} \ell W$ events
at the LHC with an integrated luminosity of ${\cal O}(100)$ inverse fb.
When combined with the BR of the W's into leptons,
this will again yield a $4 \ell + \missET$ signal comparable to that of the golden
mode $H \to ZZ \to 4 \ell$.

The FC decays $H \to \nu^\prime \bar\nu_i$ and $H \to \ell^\prime \bar\ell_i$
are expected to have a BR of ${\cal O}(10^{-4})$ if $U_{i4} \sim {\cal O}(0.1)$,
i.e., similar to the corresponding
FC decays in the quark sector. However, the leptonic
mixing angles between the 4th generation leptons and the SM light leptons
are unfortunately expected to be at best $U_{i4} \lsim {\cal O}(0.01)$
\cite{sher,PDG2008}.

Finally, let us comment on the possibility of searching for
new CP-violation effects via our Higgs decays to single
4th generation fermions. As is well known,
the extension of a 4th generation of fermions to the SM
adds two new phases to the (now 4x4) CKM, which provide
new sources of CP-violation \cite{jarlskog}.
Indeed, as was recently shown in \cite{ourFL},
the SM4 may be a natural framework for accommodating
the observed flavor and CP structure in nature, and for
addressing the CP-properties associated with
the 4th generation quarks.
Understanding the new CP-violating sector of the SM4 may also shed
light on CP-anomalies in K and b-quark systems
\cite{hou2006,soniCP,hou2005} and on baryogenesis \cite{CPbaryo}.
Effects of the new SM4 CP-violating phases can be
searched for directly in $t^\prime$ and $b^\prime$ systems
as was recently noted in \cite{gadinew,hou2009}. In particular,
studying CP-violation in these heavy 4th generation quark
systems may help
to pin down the single dominating CP-violating quantity
associated with the 4th generation quarks at high-energies \cite{paco1998}.
In this respect
our single $t^\prime$ and $b^\prime$
production channels via a heavy Higgs resonance,
$gg \to H \to t^\prime b W, ~b^\prime t W$, may be rich
in exhibiting various types of CP-asymmetries in analogy
to CP-violation in single-top production \cite{oursingletop}.

{\bf Acknowledgments:} The research of AS is supported in part by the US DOE contract \#
DE-AC0298CH10886 (BNL). SBS and GE acknowledge research support from the Technion.


\begin{thebibliography}{99}

\bibitem{sher} P.H. Frampton, P.Q. Hung and M. Sher, Phys. Rept. {\bf 330}, 263 (2000).

\bibitem{hou2009-rev}
B. Holdom, W.S. Hou, T. Hurth, M.L. Mangano, S. Sultansoy, G. Unel, talk presented
at {\it Beyond the 3rd SM generation at the LHC era workshop}, Geneva, Switzerland, Sep 2008,
arXiv:0904.4698 [hep-ph], published in PMC Phys. {\bf A3}, 4 (2009).

\bibitem{SM4proc} For older literature on the 4th generation SM, see:
Proceedings of the First (February 1987) and the Second (February 1989) International
Symposiums on the {\it fourth family of quarks and leptons}, Santa Monica, CA,
published in
Annals of the New York Academy of Sciences, 517 (1987) \& 578 (1989),
edited by D. Cline and A. Soni.

\bibitem{PDG2008} See e.g., chapter 10.7 in:
C. Amsler {\it et al.}, [Particle Data Group], Phys. Lett. {\bf B667}, 1 (2008).

\bibitem{kribs} G.D. Kribs, T. Plehn, M. Spannowsky, T.M.P. Tait,
Phys. Rev. {\bf D76}, 075016 (2007);
{\it ibid.} Nucl. Phys. Proc. Suppl. {\bf 177-178}, 241-245 (2008).

\bibitem{chan} M.S. Chanowitz, Phys. Rev. {\bf D79}, 113008 (2009).

\bibitem{vysotsky} V.A. Novikov, A.N. Rozanov and M.I. Vysotsky, arXiv:0904.4570 [hep-ph]
and references theirin.

\bibitem{bobro} M. Bobrowski, A. Lenz, J. Riedl and J. Rohrwild,
Phys. Rev. {\bf D79}, 113006 (2009).

\bibitem{hou2006} A. Arhib and W.S. Hou, JHEP {\bf 0607}, 009 (2006).

\bibitem{gadinew} G. Eilam, B. Melic and J. Trampetic, Phys. Rev. {\bf D80}, 116003 (2009).

\bibitem{soniCP} A. Soni, A.K. Alok, A. Giri, R. Mohanta, S. Nandi,
arXiv:0807.1971 [hep-ph]; {\it ibid.} arXiv:1002.0595 [hep-ph].

\bibitem{newlimittp} A. Lister (for the CDF collaboration),
talk presented at {\it 34th International Conference on High Energy Physics} (ICHEP 2008),
Philadelphia, Pennsylvania, 30 Jul - 5 Aug 2008, arXiv:0810.3349 [hep-ex].

\bibitem{newlimitbp} T. Aaltonen {\it et al.} (by the CDF Collaboration), arXiv:0912.1057 [hep-ex].

\bibitem{bardeen} B. Holdom, Phys. Rev. Lett. {\bf 57}, 2496 (1986)
[Erratum-ibid. {\bf 58}, 177 (1987);
W.A. Bardeen, C.T. Hill and M. Lindner, Phys. Rev. {\bf D41}, 1647 (1990);
C. Hill, M. Luty and E.A. Paschos, Phys. Rev. {\bf D43}, 3011 (1991);
P.Q. Hung and G. Isidori Phys. Lett. {\bf B402}, 122 (1997).

\bibitem{soni-composite} For another early work where the relation $m_H \sim \sqrt{2} m_Q$
is in a composite model for the Higgs, see,
J. Carpenter, R. Norton, S. Siegemund-Broka and A. Soni, Phys. Rev. Lett. {\bf 65}, 153 (1990).

\bibitem{recentcomposite} P.Q. Hung and Chi Xiong, arXiv:0911.3890 [hep-ph];
{\it ibid.} arXiv:0911.3892 [hep-ph].

\bibitem{recentcomposite2} M. Hashimoto and V.A. Miransky, arXiv:0912.4453 [hep-ph].

\bibitem{gunion} J.F. Gunion, D.W. McKay and H. Pois,
Phys. Lett. {\bf B334}, 339 (1994); {\it ibid.} Phys. Rev. {\bf D53}, 1616 (1996).

\bibitem{cont2} H.-J. He, N. Polonsky and S. Su, Phys. Rev. {\bf D64}, 053004 (2001).

\bibitem{kfactor} See e.g.,
M. Gomez-Bock, M. Mondragon, M. Muhlleitner, M. Spira and P.M. Zerwas,
Lectures given at {\it 4th CERN-CLAF School of High-Energy Physics},
Vina Del Mar, Valparaiso Region, Chile, Feb 2007, arXiv:0712.2419 [hep-ph] and references theirin,
published in Vina del Mar, High energy physics, 177 (2007).

\bibitem{widtheffects} G. Altarelli, L. Conti and V. Lubicz, Phys. Lett. {\bf B502}, 125 (2001);
G. Mahlon and S. J. Parke, Phys. Lett. {\bf B347}, 394 (1995);
T. Muta, R. Najima and S. Wakaizumi, Mod. Phys. Lett. {\bf A1}, 203 (1986);
G. Calderon and G. Lopez Castro, arXiv: hep-ph/0108088;
V.I. Kuksa, arXiv: hep-ph/0404281;
V.I. Kuksa, Phys. Lett. {\bf B633}, 545 (2006), Erratum-ibid. {\bf B664}, 315 (2008);
S. Bar-Shalom, G. Eilam, M. Frank, I. Turan, Phys.Rev. {\bf D72}, 055018 (2005).

\bibitem{calchep} CalcHEP - a package for calculation of Feynman diagrams and
integration over multi-particle phase space, by A. Pukhov, A. Belyaev and N. Christensen,\\
http://theory.sinp.msu.ru/$\sim$pukhov/calchep.html;
see also, A. Pukhov {\it et al.}, hep-ph/9908288 and A. Pukhov, hep-ph/0412191.

\bibitem{HtotcSM} See e.g.,
I. Baum, G. Eilam and S. Bar-Shalom, Phys. Rev. {\bf D77}, 113008 (2008).


\bibitem{behar} S. Bejar, J. Guasch and J. Sola, Nucl. Phys. {\bf B675}, 270 (2003).


\bibitem{GSH} G. Eilam, B. Haeri and A. Soni, Phys. Rev. Lett {\bf 62}, 719 (1989);
{\it ibid.} Phys. Rev. {\bf D41}, 875 (1990).

\bibitem{bpdecays} W.-S. Hou and R.G. Stuart, Phys. Rev. Lett. {\bf 62}, 617 (1989);
{\it ibid.} Nucl. Phys. {\bf B320}, 277 (1989);
{\it ibid.} Phys. Lett. {\bf B233}, 485 (1989);
{\it ibid.} Nucl. Phys. {\bf B349}, 91 (1991);
{\it ibid.} Phys. Rev. {\bf D43}, 3669 (1991);
M. Sher, Phys. Rev. {\bf D61}, 057303 (2000);
A. Arhrib and W.-S. Hou, Phys. Rev. {\bf D64} 073016 (2001).

\bibitem{unel} T.C.-Donszelmann, M.K. Unel, V.E. Ozcan, S. Sultansoy and G. Unel,
JHEP {\bf 0810}, 074 (2008);
K. Belotsky, D. Fargion, M. Khlopov, R. Konoplich, K. Shibaev, Phys. Rev. {\bf D68}, 054027 (2003).

\bibitem{jarlskog} C. Jarlskog, Phys. Rev. {\bf D36}, 2128 (1987).

\bibitem{hou2005} W.-S. Hou, M. Nagashima and A. Soddu,
Phys. Rev. {\bf D72}, 115007 (2005);
W.-S. Hou, M. Nagashima and A. Soddu,
Phys. Rev. Lett. {\bf 95}, 141601 (2005);
W.-S. Hou, H.-nan Li, S. Mishima and M. Nagashima,
Phys. Rev. Lett. {\bf 98}, 131801 (2007).


\bibitem{CPbaryo} W.S. Hou, Chin. J. Phys. {\bf 47}, 134 (2009),
arXiv:0803.1234 [hep-ph];
W.S. Hou, talk given at {\it 34th International Conference on High Energy Physics (ICHEP 2008)},
Philadelphia, Pennsylvania, Jul 2008, arXiv:0810.3396 [hep-ph]; see, however,
R. Fok and G.D. Kribs, Phys. Rev. {\bf D78}, 075023 (2008).

\bibitem{ourFL} S. Bar-Shalom, D. Oaknin and A. Soni, Phys. Rev. {\bf D80}, 015011 (2009).

\bibitem{hou2009} A. Arhrib and W.-S. Hou, Phys. Rev. {\bf D80}, 076005 (2009).


\bibitem{paco1998} F. del Aguila, J.A. Aguilar-Saavedra and G.C. Branco,
Nucl. Phys. {\bf B510}, 39 (1998).


\bibitem{oursingletop} D. Atwood, S. Bar-Shalom, G. Eilam and A. Soni, Phys. Rev. {\bf D54}, 5412 (1996);
{\it ibid.}, Phys. Rept. {\bf 347}, 1 (2001).

\end{thebibliography}
\end{document}